\documentclass[
preprint,
showpacs,amsmath,amssymb,aps,pra
]{revtex4-1}

\usepackage{graphicx}
\usepackage{dcolumn}
\usepackage{bm}

\newcommand{\eref}[1]{(\ref{#1})}
\newcommand{\etal}{{\it et al.}}

\begin{document}

\title{Quenched decoherence in qubit dynamics due to strong amplitude-damping noise}

\author{Shin-Tza Wu}
\email{phystw@gmail.com} \affiliation{Department of Physics,
National Chung Cheng University, Chiayi 621, Taiwan}

\date{\today}

\begin{abstract}
We study non-perturbatively the time evolution of a qubit subject to
amplitude-damping noise. We show that at strong coupling the qubit
decoherence can be quenched owing to large environment feedback,
such that the qubit can evolve coherently even in the long-time
limit. As an application, we show that for a quantum channel that
consists of two independent qubits subject to uncorrelated local
amplitude-damping noises, it can maintain at strong coupling finite
entanglement and better than classical teleportation fidelity at
long times.
\end{abstract}


\maketitle

{\em Introduction}--The development of quantum information sciences has promised to bring computing and information processing
into new horizons \cite{NC}. However, these have been enormously hindered by the
inevitable coupling between the quantum bits (qubits) and their environments which disrupts the coherence
in the qubits and impedes the designated quantum tasks. In order to combat against these unwanted couplings,
different approaches have been developed since the early days of quantum information sciences.
One common approach is to apply error-correcting codes that can help identify and correct errors in
the qubits caused by the environment noise \cite{Djo}.
In more sophisticated schemes, one seeks to avoid qubit errors by encoding the (logical) qubits in
decoherence-free subspaces, namely the portion of the state space that is immune to environment noise
due to symmetries in the coupling. Such error avoidance can also be achieved dynamically by
creating decoherence-free subspaces through controlled pulses applied over the qubits \cite{Lid}.
In recent years, there has been intense interest in approaches that attempt to
encode qubits in nonlocal manners by exploiting the topological structure of
the qubit configuration space, so that the qubits can be robust against local
environment noise \cite{Top}.

In this work, we propose a scheme for quenching qubit decoherence
when amplitude-damping noise is present. Instead of avoiding the
environment coupling, we propose to couple the qubit {\em strongly}
to the environment so that a coherent dynamics can be attained. As
we will explain, this counter-intuitive result arises from the large
environment feedback over the qubit when the coupling is strong,
which keep the qubit decoherence frozen after a short time. The
qubit will then evolve coherently and reach a steady state
subsequently. Similar results have previously been demonstrated for
damped quantum harmonic oscillators \cite{Xion,WM12} and applied to
the dynamics of a nanocavity coupled to a structured waveguide
\cite{Lei}. Here we demonstrate the effect for qubit dynamics and
apply it to a quantum channel that consists of two entangled qubits
subject to local amplitude-damping noises. As we will show, the
strong coupling effect can help preserve a robust quantum channel at
long times.

{\em Formulation}--Let us start by considering a single qubit interacting with an
environment that consists of a wide spectrum of harmonic oscillators.
The Hamiltonian for the total system thus reads (we take $\hbar=1$ throughout)
\begin{eqnarray}
H =  \frac{\omega_0}{2} \sigma_z + \sum_k \omega_k b_k^\dagger b_k + H_I \, .
\label{H_tot}
\end{eqnarray}
Here $\omega_0$ is the energy separation between the qubit levels
$|0\rangle$ and $|1\rangle$ (which have energies $\pm
\frac{\omega_0}{2}$, respectively), $\sigma_z$ is the third Pauli
matrix, and $b_k$ the annihilation operator for the environment
mode which has frequency $\omega_k$. The last term $H_I$ in \eref{H_tot} describes
the interaction between the qubit and its environment. For amplitude-damping channels,
the qubit interacts with the environment modes by exchanging energies, so that \cite{NC,BP}
\begin{eqnarray}
H_I = \sum_k \left( g_k \sigma_- b_k^\dagger + g_k^* \sigma_+ b_k \right)
\, ,
\label{HI}
\end{eqnarray}
where $g_k$ are the coupling amplitudes and $\sigma_\pm$ the
raising/lowering operators for the qubit levels. We shall assume that the environment
starts initially in the vacuum state at zero temperature. In this case, the
qubit dynamics under the influence of the coupling \eref{HI} can be solved exactly \cite{Berry}.
One can express the time evolution for the reduced density matrix $\rho_s$ for
the qubit as an operator sum \cite{Kraus}
\begin{eqnarray}
\rho_s(t) = \sum_{i=1}^2 E_i \rho_s(0) E_i^\dagger \, ,
\label{Kr_1qb}
\end{eqnarray}
where $\rho_s(0)$ is the initial density matrix for the qubit and $E_i$ are the operation elements for the time
evolution of the qubit, which account for the entire environment effects over the qubit and satisfy
$\sum_{i=1}^2 E_i^\dagger E_i=1$. Explicitly, in the basis $\{|0\rangle,|1\rangle\}$, we have
\begin{eqnarray}
E_1 = \left(
           \begin{array}{cc}
                 p(t)  &  0 \\
                 0  &  1
           \end{array}
      \right) \, ,
      \quad
E_2 = \left(
           \begin{array}{cc}
                 0  &  0 \\
                 q(t)  &  0
           \end{array}
      \right) \, ,
\label{E_1qb}
\end{eqnarray}
where $q(t)\equiv\sqrt{1-|p(t)|^2}$ and $p(t)$ satisfies the equation of motion \cite{Berry}
\begin{eqnarray}
\frac{d}{d t}p(t) = - \int_0^t d\tau f(t-\tau) p(\tau)
\label{eom}
\end{eqnarray}
subject to the initial condition $p(0)=1$. Note that here we are using the interaction picture
and have introduced the noise correlation function
\begin{eqnarray}
f(t-\tau) &=& \sum_{k} |g_k|^2 \, e^{i(\omega_0-\omega_k)(t-\tau)}
\nonumber\\
&=& \int_0^\infty d\omega\, J(\omega)\, e^{i(\omega_0-\omega)(t-\tau)}
\label{fJ}
\end{eqnarray}
with $J(\omega)$ the spectral function for the environment coupling \cite{BP}.
To study the qubit dynamics, we will examine the time evolution of the Bloch vector
$\vec{r}(t)=\mbox{Tr}\{\rho_s(t)\vec{\sigma}\}$ (with $\vec{\sigma}$ the Pauli matrices)
for given initial qubit state. Making use of \eref{Kr_1qb} and \eref{E_1qb},
one can obtain easily
\begin{eqnarray}
r_x(t)&=& 2 \mbox{Re}\{p(t)\rho_{s,12}(0)\} \, ,
\nonumber\\
r_y(t)&=&-2 \mbox{Im}\{p(t)\rho_{s,12}(0)\} \, ,
\nonumber\\
r_z(t)&=& 2 |p(t)|^2\rho_{s,11}(0) - 1 \, ,
\label{Bv}
\end{eqnarray}
where $\rho_{s,ij}(0)$ indicate the $ij$ element of the initial density matrix.
The qubit dynamics is thus governed entirely by the function $p(t)$, which can be
solved from \eref{eom} for given spectral function $J(\omega)$.
For later purposes, we note that an exact master equation for the qubit evolution
in the interaction picture can also be derived based on \eref{Kr_1qb} and \eref{E_1qb} \cite{Berry}
\begin{eqnarray}
\frac{d\rho_s(t)}{dt} &=& -i S(t) [\sigma_+\sigma_-,\rho_s(t)]
\nonumber\\ &+&\!
\gamma(t)\!\left(2\sigma_-\rho_s(t)\sigma_+\!\!-\!\sigma_+\sigma_-\rho_s(t)\!-\!\rho_s(t)\sigma_+\sigma_-\!\right),
\label{me}
\end{eqnarray}
where $S(t)=-{\rm Im}(\dot{p}/p)$ is a time-dependent frequency shift and $\gamma(t)=-{\rm Re}(\dot{p}/p)$
a time-changing decay rate. As is evident from \eref{me}, when $\gamma(t)$ becomes negative,
it indicates an energy back flow from the environment to the qubit \cite{Xion}. This feedback effect will
be crucial for quenching qubit decoherence that we shall discuss below.

\begin{figure}
\includegraphics*[width=50mm]{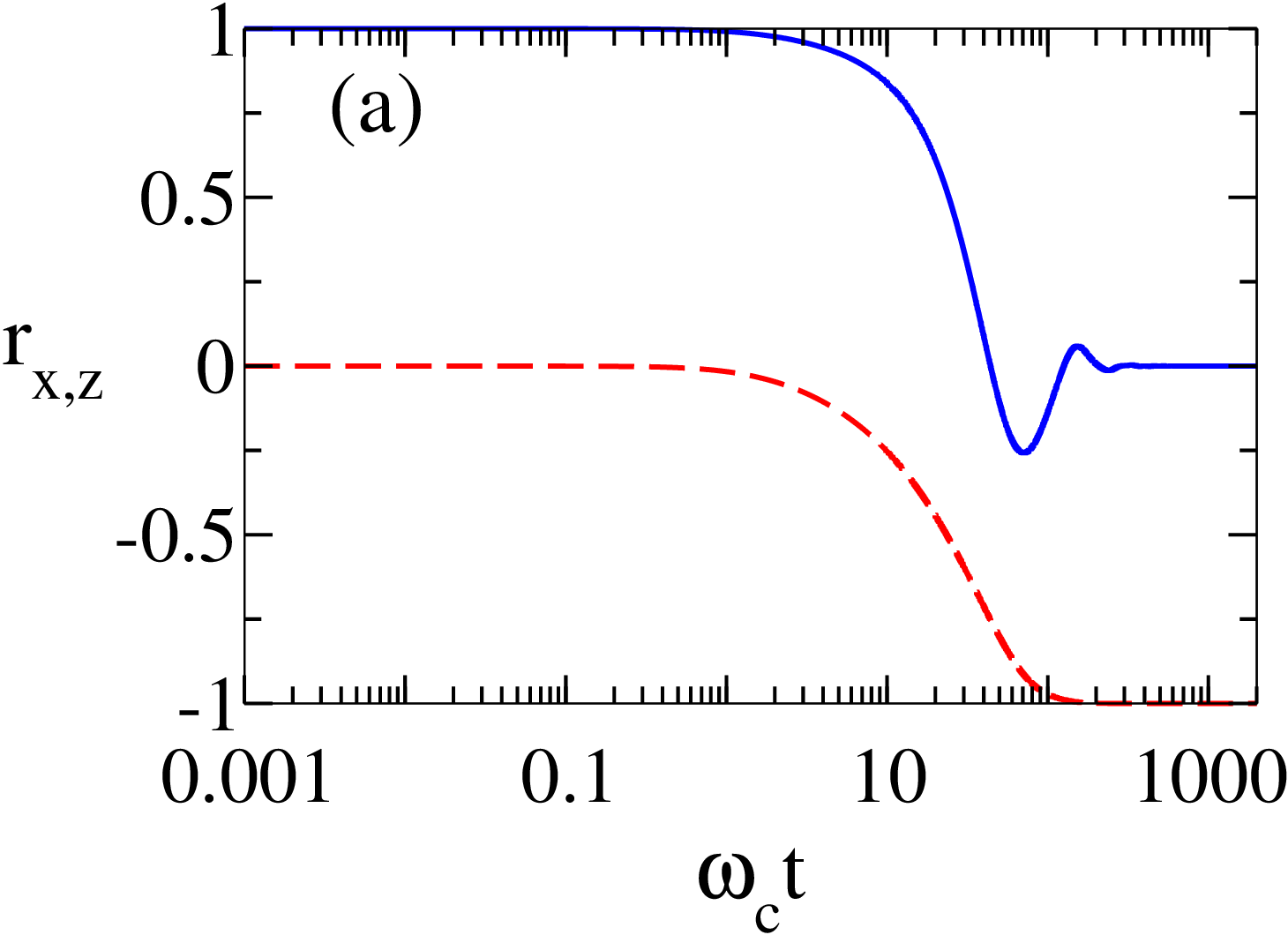}
\includegraphics*[width=50mm]{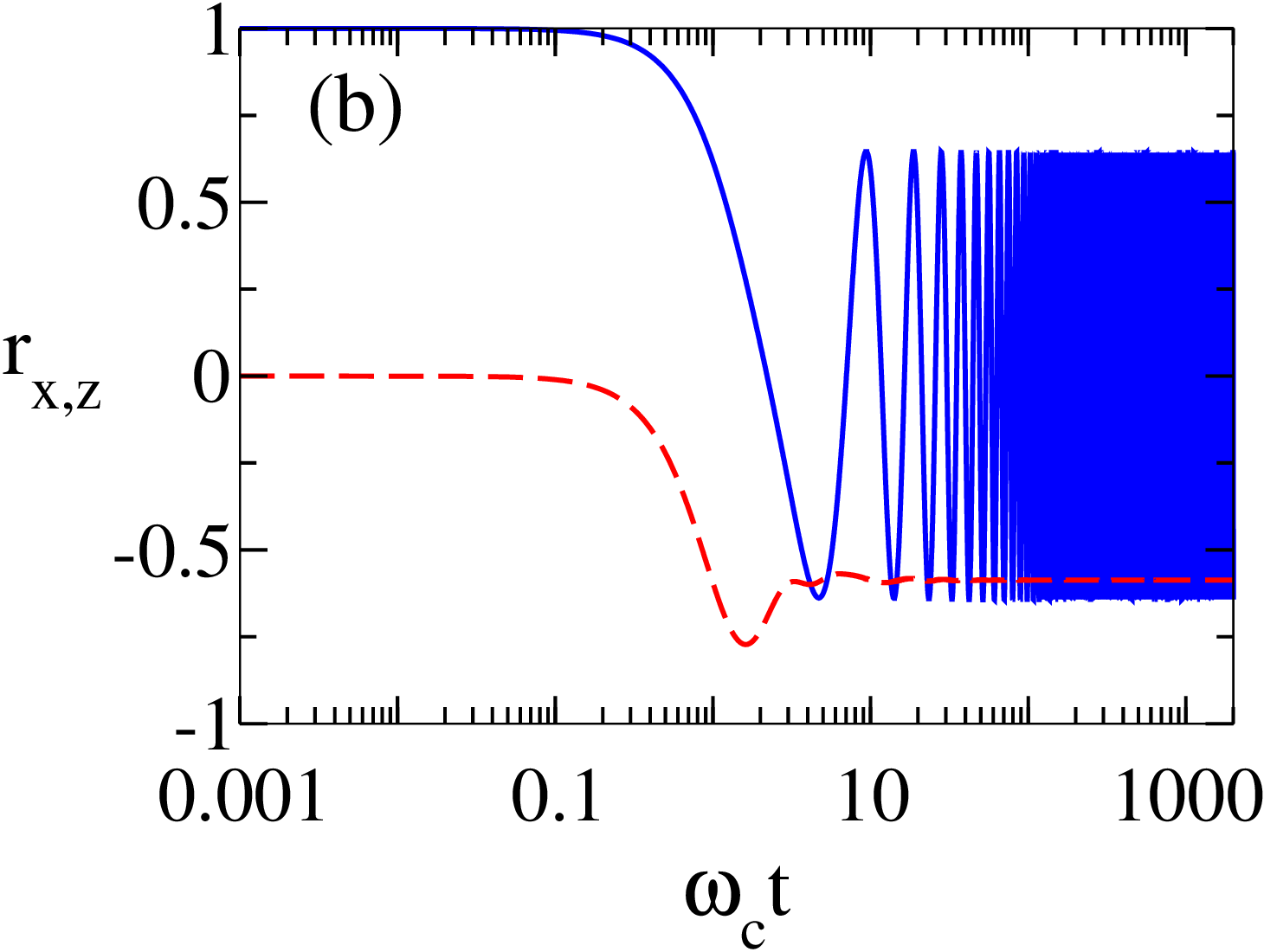}
\caption{Time evolution of the $x$-component (blue solid curves) and
$z$-component (red dashed curves) of the Bloch vector \eref{Bv}
for the cat state \eref{cat} at (a) weak coupling ($\eta_0=0.01$)
and (b) strong coupling ($\eta_0=0.5$). Note that for all figures in
this paper, we plot the time axis in log scale.
\label{fig:SxSz}}
\end{figure}

\begin{figure}
\includegraphics*[width=50mm]{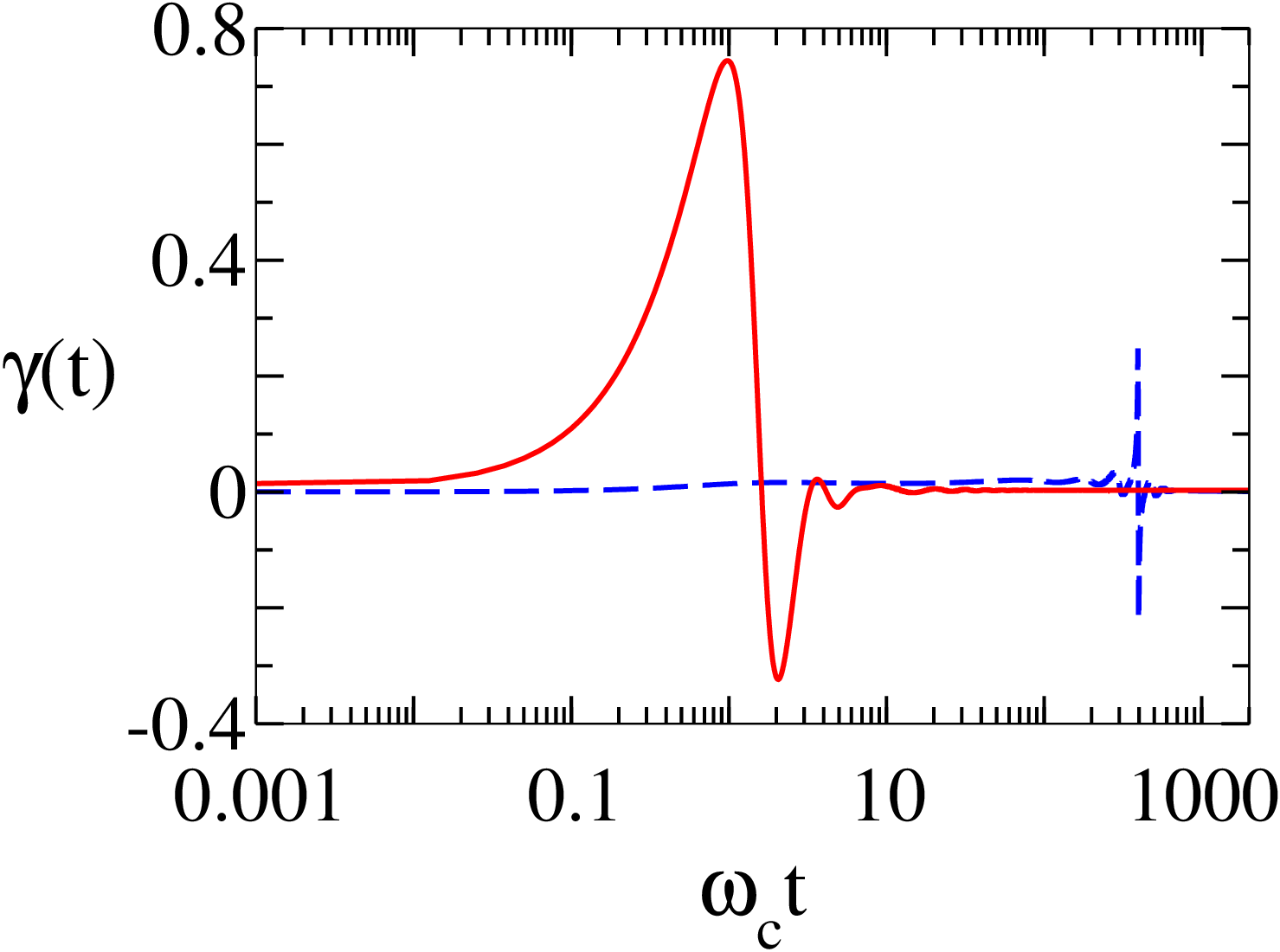}
\caption{Time dependence for the decay rate $\gamma(t)$ in the master equation \eref{me}
for weak coupling ($\eta_0=0.01$; blue dashed curve) and strong coupling ($\eta_0=0.5$; red solid curve).
\label{fig:rate}}
\end{figure}

Let us consider spectral functions that have power-law frequency dependence and an exponential cutoff \cite{yw13}
\begin{eqnarray}
J(\omega)=\eta_s \omega \left(\frac{\omega}{\omega_c}\right)^{s-1}
\exp\left(-\frac{\omega}{\omega_c}\right) \, ,
\label{Jw}
\end{eqnarray}
where $\eta_s=\eta_0(e/s)^s$ with $\eta_0$ the coupling strength, and $\omega_c$ is
the cutoff frequency. According to the power $s$, the coupling induced by $J(\omega)$ is usually
categorized as being sub-Ohmic ($0<s<1$), Ohmic ($s=1$), and super Ohmic ($s>1$) \cite{Weiss}.
As a demonstration, let us consider a qubit initially in the cat state
\begin{eqnarray}
|\psi\rangle = \frac{|0\rangle + |1\rangle}{\sqrt{2}} \, .
\label{cat}
\end{eqnarray}
For the spectral function \eref{Jw}, one can solve $p(t)$ numerically from the equation of
motion \eref{eom} using Laplace transform \cite{yw13}. The qubit dynamics then follows from
Eqs.~\eref{Kr_1qb} and \eref{E_1qb}. Figure \ref{fig:SxSz} illustrates for the cat state \eref{cat}
the time evolution of the $x$ and $z$ components of the Bloch vector \eref{Bv} when the
environment coupling is sub-Ohmic with power $s=1/2$. Here and in all subsequent plots, we
take $\omega_0=0.1\omega_c$ for the qubit. We see that at weak coupling ($\eta_0=0.01$) the qubit decays
to the lower level $|1\rangle$ asymptotically and its coherence $r_x$ vanishes completely at long times.
At strong coupling ($\eta_0=0.5$), however, $r_z$ attains a non-zero stationary value while $r_x$
oscillates coherently even at long times. Namely, after decoherence during the first few periods,
the qubit coherence can be preserved even at long times when the qubit is coupled strongly
to the amplitude-damping noise. To see the reason, we calculate in
Fig.~\ref{fig:rate} the decay rate $\gamma(t)$ in the exact master equation \eref{me}.
We observe that at weak coupling, the decay rate remains finite
as long as the Bloch vector has not yet fully relaxed to $r_x=0$ and $r_z=-1$. Although a
late-stage feedback (i.e.~sign change in $\gamma(t)$) occurs near $\omega_ct\simeq 396.4$,
its duration is too short to recover the qubit coherence. At strong coupling, however, although the decay
rate has large values initially (due to the strong coupling), environment feedback brings it
negative at an early stage of the time evolution. The decay rate then settles to zero very
quickly, leading to the coherent dynamics seen in Fig.~\ref{fig:SxSz} (b) \cite{note}. Since $\gamma(t)$
depends only on $p(t)$, these results are fairly general and independent of the initial qubit state.
Moreover, as noted earlier, as the qubit dynamics is controlled chiefly by $p(t)$, the quenched
qubit decoherence is therefore a general strong-coupling effect that does not depend on the initial
qubit state.

{\em Two qubits}--Let us now turn to the problem of a pair of independent qubits
that are subject to local amplitude-damping noises, with each qubit coupled to its
local environment according to \eref{HI}.
Since the pair of qubits are not interacting with each other and
their environment noises are not correlated, the two-qubit dynamics
can be constructed from the single-qubit formulas utilizing the Kraus
formulation \cite{YE04}.  The reduced density matrix $\rho$
for the two qubits thus evolves according to
\begin{eqnarray}
\rho(t) = \sum_{i,j=1}^2 \left(E_i^A\otimes E_j^B\right) \rho(0) \left(E_i^A\otimes E_j^B \right)^\dagger \, ,
\label{Kr_2qb}
\end{eqnarray}
where the superscripts $A$, $B$ are the qubit labels and $E_i^\alpha$
(with $\alpha=A,B$) are the operation elements in \eref{E_1qb}
with $p_\alpha$ satisfying the equation of motion \eref{eom} for the respective qubits.

As an application, let us consider a quantum channel made of two qubits
initially in the Bell state
\begin{eqnarray}
|\Phi^+\rangle = \frac{|00\rangle+|11\rangle}{\sqrt{2}} \,.
\label{bell}
\end{eqnarray}
When each qubit in the quantum channel is subject to
amplitude-damping noise, the quantum channel would degrade with
time. As a measure for the quality of the quantum channel, we shall
examine the time evolution of its concurrence \cite {Woot} and
maximum teleportation fidelity \cite{Horod}. Since the initial
density matrix $\rho(0)=|\Phi^+\rangle\langle\Phi^+|$ has an X-form
\cite{YE_X} in the basis
$\{|00\rangle,|01\rangle,|10\rangle,|11\rangle\}$, under the time
evolution \eref{Kr_2qb}, $\rho(t)$ will retain the form \cite{note2}
\begin{eqnarray}
\rho(t) = \left(
                \begin{array}{cccc}
                      \rho_{11}(t) &     0        &      0       & \rho_{14}(t) \\
                          0        & \rho_{22}(t) & \rho_{23}(t) &     0     \\
                          0        & \rho_{32}(t) & \rho_{33}(t) &     0     \\
                      \rho_{41}(t) &     0        &      0       & \rho_{44}(t)
                \end{array}
          \right) \, .
\label{rho_X}
\end{eqnarray}
The concurrence and the maximum teleportation fidelity for the quantum channel
\eref{rho_X} can thus be obtained straightforwardly. We find the concurrence
\begin{eqnarray}
C(t) = 2 \max \Big\{\, 0 , & |\rho_{14}(t)|-\sqrt{\rho_{22}(t)\rho_{33}(t)} \,,
\nonumber\\
& |\rho_{23}(t)|-\sqrt{\rho_{11}(t)\rho_{44}(t)} \, \Big\}
\label{Cx}
\end{eqnarray}
and the maximum teleportation fidelity
\begin{eqnarray}
F(t)=\frac{1}{3}\bigg[1+\max\{&\rho_{11}(t)+\rho_{44}(t)+2|\rho_{14}(t)|\,,
\nonumber\\
&\rho_{22}(t)+\rho_{33}(t)+2|\rho_{23}(t)|\}\bigg] \, .
\label{Fx}
\end{eqnarray}
For simplicity, let us suppose the two qubits are completely identical,
so that they have the same level separation $\omega_0$ and the same coupling to
their local environments. One can thus drop the qubit labels $A,B$ in \eref{Kr_2qb}
and, upon using explicit expressions for the operation elements \eref{E_1qb},
obtain from \eref{Cx} and \eref{Fx}
\begin{eqnarray}
C(t)&=&|p(t)|^4 \, ,
\nonumber \\
F(t)&=&\frac{1}{3} \bigg[ 1+\max\{1+|p(t)|^4, |p(t)|^2 q(t)^2\}\bigg] \, .
\label{CtFt}
\end{eqnarray}

\begin{figure}
\includegraphics*[width=50mm]{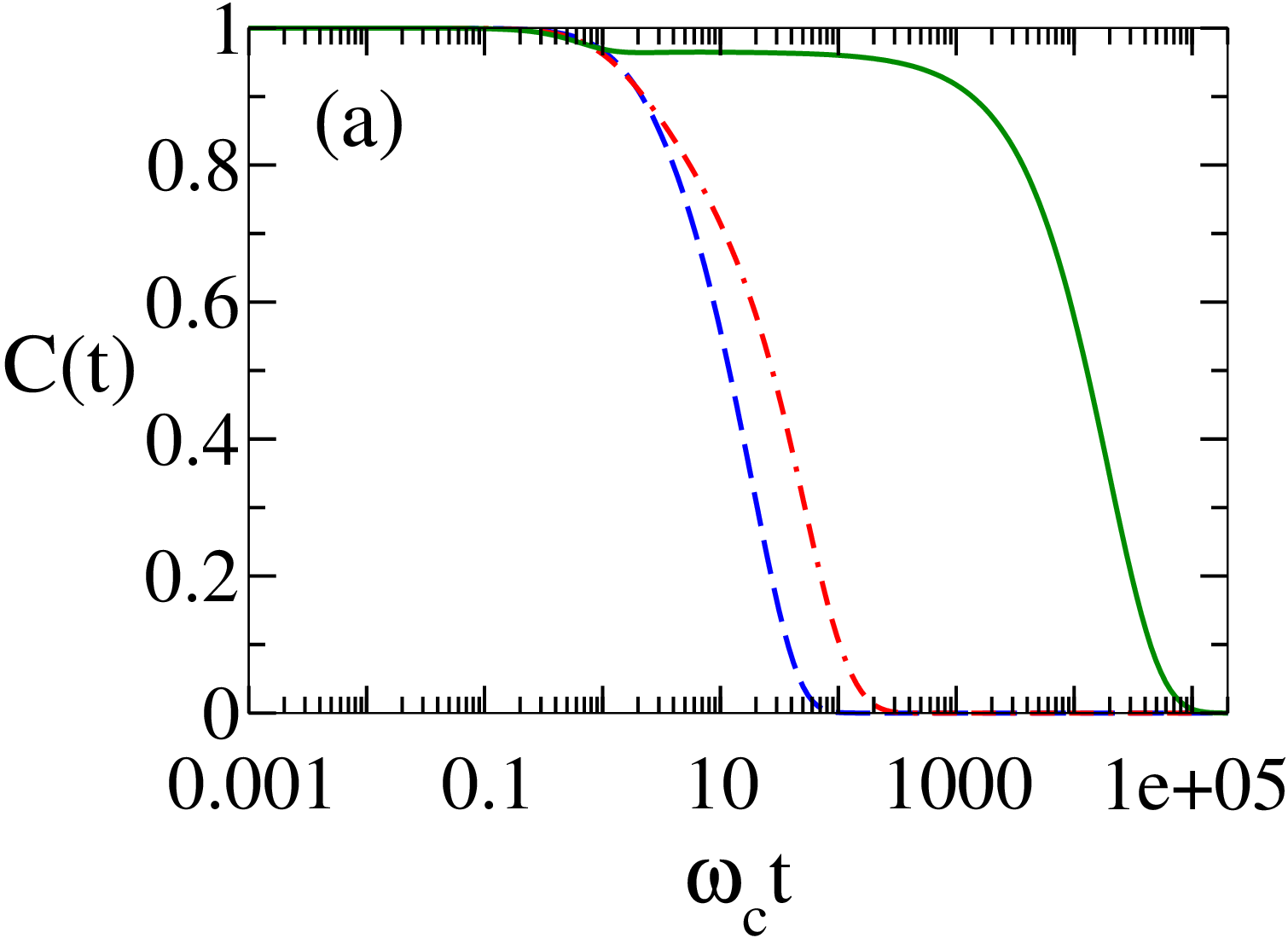}
\includegraphics*[width=50mm]{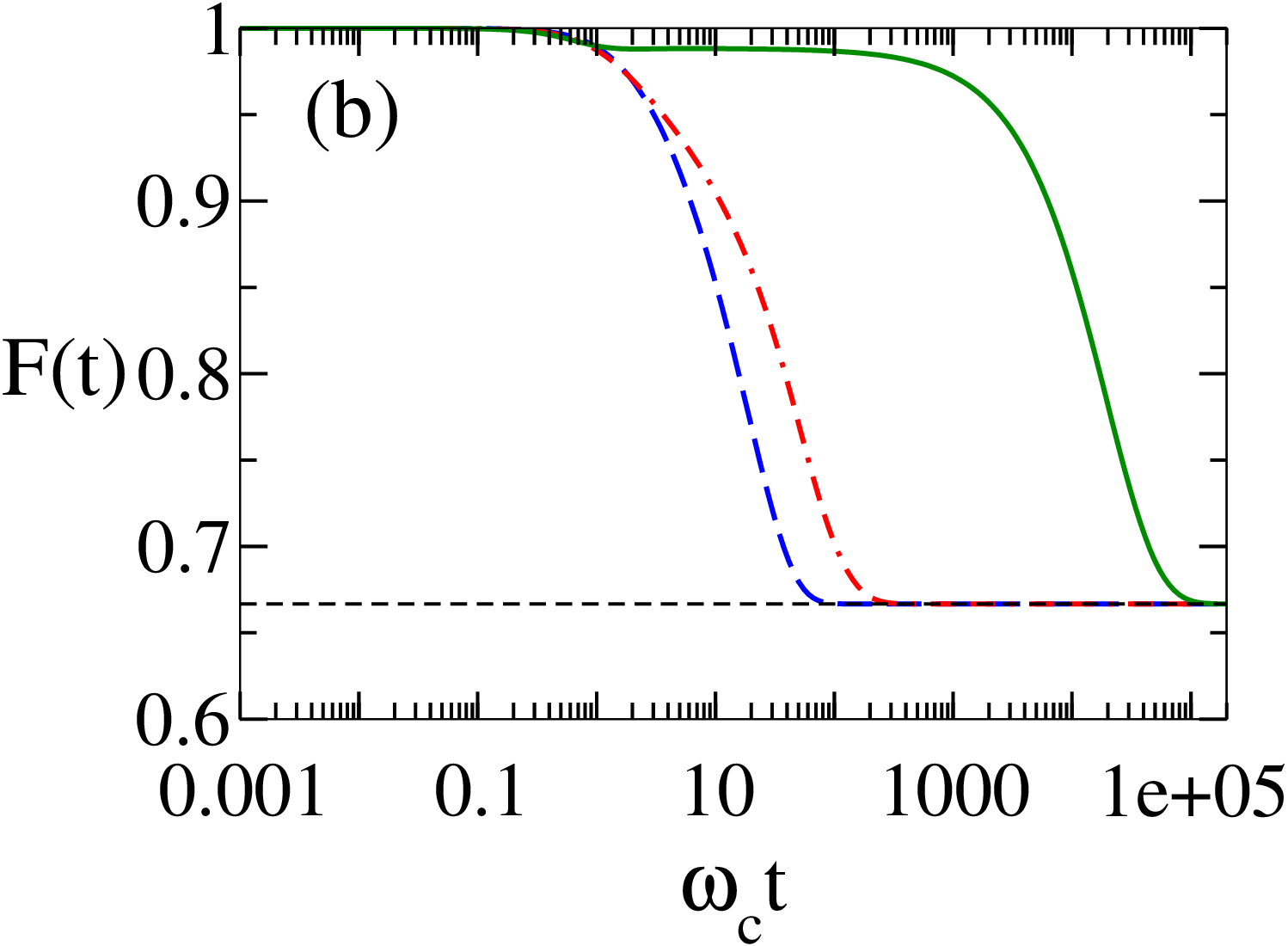}
\caption{Time evolution for the (a) concurrence and (b) maximum teleportation fidelity for
the Bell state \eref{bell} at weak environment coupling ($\eta_0=0.01$) when the spectral function is sub-Ohmic
($s=1/2\,$; blue dashed curves), Ohmic ($s=1\,$; red dot-dashed curves), and super-Ohmic ($s=3\,$; green solid curves).
The horizontal dashed line in (b) indicates the classical limit $F=\frac{2}{3}$.
\label{fig:eta001}}
\end{figure}

\begin{figure}
\includegraphics*[width=50mm]{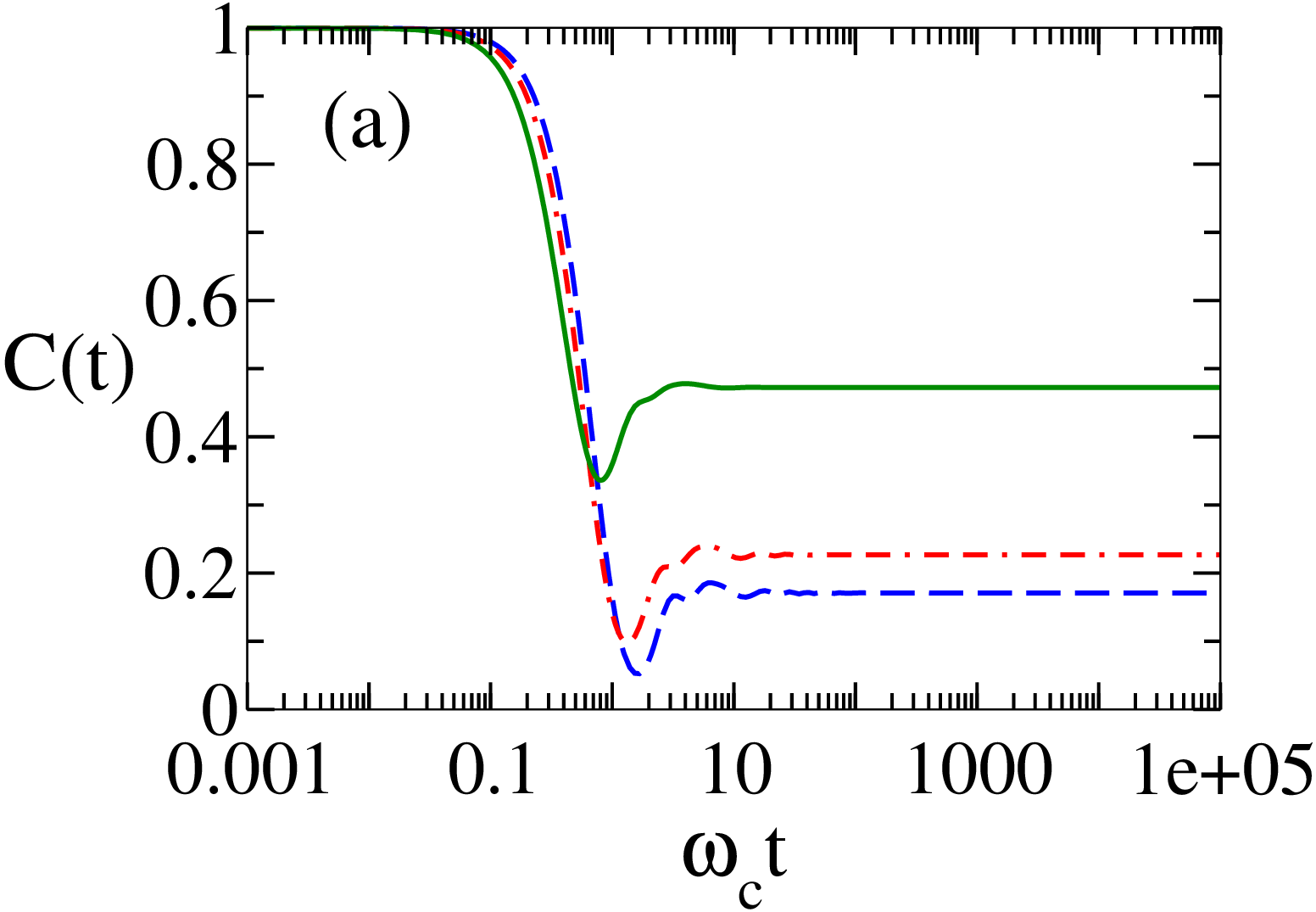}
\includegraphics*[width=50mm]{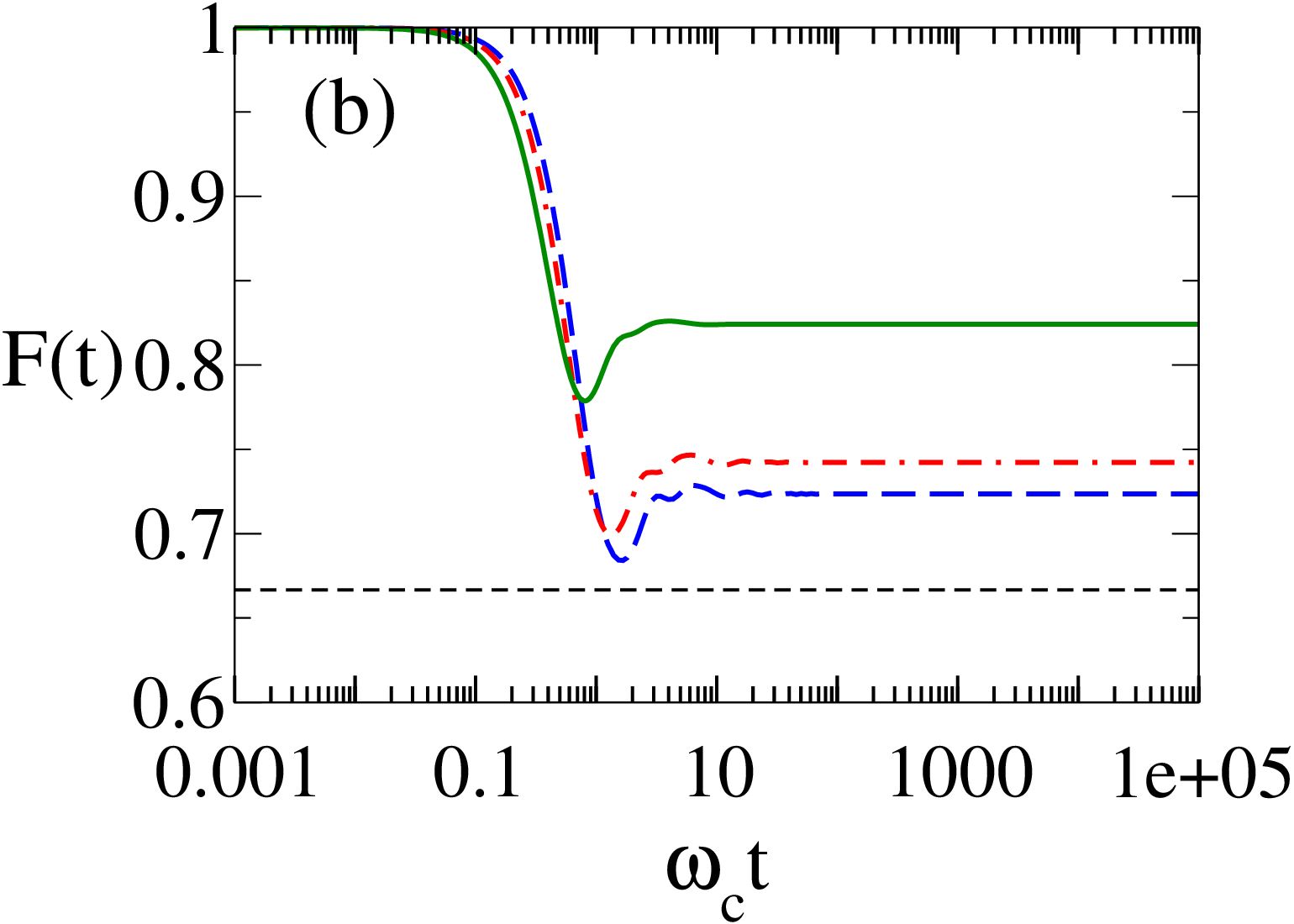}
\caption{The same as Fig.~\ref{fig:eta001} but for strong coupling ($\eta_0=0.5$).
\label{fig:eta05}}
\end{figure}

For explicit calculations, as previously, we take $\omega_0=0.1\omega_c$ for the qubits and consider the spectral
function \eref{Jw} for the environment coupling. Feeding $p(t)$ solved from \eref{eom} into Eq.~\eref{CtFt},
one can obtain the time evolution for the concurrence and the teleportation fidelity for the
quantum channel. Figure \ref{fig:eta001} shows our results for the coupling strength $\eta_0=0.01$.
We see that for sub-Ohmic, Ohmic, and super-Ohmic couplings, the concurrence decay to zero monotonically and
the teleportation fidelity reduce to the classical value $F=2/3$ asymptotically \cite{Pope}. Since for given $\eta_0$ and $\omega_c$
the spectral function \eref{Jw} would have the same peak strength $\eta_0\omega_c$ for all $s>0$. For different
power $s$ in $J(\omega)$, the influence of the coupling over the qubits can therefore be compared according to
the detuning of the peak frequency
(at $\omega=s\omega_c$) from the qubit frequency $\omega_0$ \cite{yw13}. For the results in Fig.~\ref{fig:eta001},
since the sub-Ohmic case has the smallest detuning among the three while the super-Ohmic one has the largest,
the former would induce the strongest decoherence in the qubits while the latter the weakest.

At stronger coupling $\eta_0=0.5$, Fig.~\ref{fig:eta05} shows that
for all three types of coupling the concurrence can maintain finite
values even at long times and the teleportation fidelity always stay
above the classical value. In particular, the (most detuned)
super-Ohmic case has large stationary values $C\simeq 0.472$ and
$F\simeq 0.824$. As is clear from \eref{CtFt}, like single-qubit
case, these results are connected with the undamped time evolution
of $p(t)$ at strong coupling. It is therefore the large feedback due
to strong environment coupling that result in the robust quantum
channel in the long-time limit. Although stationary entanglement
evolution has previously been observed in other systems
[\onlinecite{Daj07}--\onlinecite{Hue}, \onlinecite{yw13}], here the
effect is entirely intrinsic \cite{Daj07,Daj08}, in that the quantum
channel can survive the environment noise in the long-time limit
without any external driving \cite{Ang,Li}, mutual interaction
between the qubits \cite{Hue}, or error-correcting codes
\cite{yw13}.

{\em Discussions}--The model Hamiltonians \eref{H_tot} and \eref{HI} considered in this paper are quite general. They can
describe, for instance, a two-level atom inside a cavity, an exciton in a potential well, a Cooper pair box
coupled to a transmission line, and so forth. Our conclusions should thus be applicable to these systems. As with damped
quantum harmonic oscillators \cite{Xion}, the quenched decoherence in the qubit
dynamics is a consequence of the non-dissipative $p(t)$ evolution. The existence of such solutions for \eref{eom}
depends crucially on the form of the spectral function \cite{WM12}. As was shown in Ref.~\cite{wu12},
for Lorentzian spectral functions, $p(t)$ would vanish in the long-time limit except for
the trivial case of vanishing coupling and single-mode environment interaction. For the spectral function \eref{Jw},
as illustrated in Fig.~\ref{fig:eta05}, non-dissipative $p(t)$ can occur for different power $s$ if the
coupling is strong enough. Therefore, experimental engineering for the spectral function \eref{Jw}
need not be stringent over the power $s$ (although it would certainly affect the
``quality" of the steady state).

Although an entirely decoherence-free qubit dynamics was not achieved in the present scheme, the
``frozen" qubit coherence can pave ways for further applications. For instance, since the strong-coupling
effect allows the qubit to evolve coherently for arbitrarily long durations (see Fig.~\ref{fig:SxSz} (b)),
it therefore provides a means of storing qubits. A major challenge here is clearly how to restore the impaired
qubit coherence and enhance the storage fidelity at readout. For this purpose, it may be useful to incorporate
purification protocols \cite{Ben,Bri} or error-correcting codes \cite{Sho,Fle} into the scheme. The same
applies also to the robust quantum channel discussed earlier. We are presently investigating such
possibilities.

In conclusion, based a non-perturbative approach we have studied the
time evolution of qubits subject to amplitude-damping noises. We
have shown that at strong coupling, the qubit can maintain coherence
at long times due to large environment feedback. We have also
demonstrated that for two qubits that undergo local decoherence due
to amplitude-damping noises, they can preserve at long times finite
entanglement and better than classical teleportation fidelity.

I would like to thank Prof. Dian-Jiun Han for valuable discussions. I am also
grateful to Drs. R. Lo Franco and S. Paraoanu for bringing to my attention relevant references.
This work is supported by NSC of Taiwan through grant no.~NSC 99-2112-M-194-009 -MY3;
it is also partly supported by the Center for Theoretical
Sciences, Taiwan.

{\em Note added in proof.} While this paper was under review,
I became aware of Ref.~\cite{Liu}, which discusses a quantum
phase-transition that underlies the appearance and disappearance of
coherent steady-states for a two-level system subject to
environmental noise when the coupling strength is varied. I would
like to thank Prof.~Jun-Hong An for bringing this work to my notice.



\begin{thebibliography}{99}
\bibitem{note}
In Ref.~\onlinecite{Xion}, similar observation was made for damped quantum
harmonic oscillators strongly coupled to harmonic baths.
\bibitem{note2}
In fact, for the initial state \eref{cat} we have $\rho_{23}(t)=0$
always. Here we retain the general X-form in order that the subsequent
equations \eref{Cx} and \eref{Fx} will have wider applicability.
\end{thebibliography}
\end{document}